\documentclass[print]{revtex4}

\usepackage{amsmath,amssymb}
\usepackage{graphicx}
\newcommand{\lvec}[1]{|#1\!\!>}

\draft

\begin{document}

\title{\large \bf Solitons Formed by Dark-State Polaritons in
Electromagnetic Induced Transparency}
\author{Xiong-Jun Liu$^{a,b}$\footnote{Electronic address: x.j.liu@eyou.com}, Hui Jing$^{c}$ and Mo-Lin Ge$^{a,b}$}
\affiliation{a. Theoretical Physics Division, Nankai Institute of
Mathematics,Nankai University, Tianjin 300071, P.R.China\\
 b. Liuhui Center for Applied Mathematics, Nankai
University and Tianjin University, Tianjin 300071, P.R.China\\
 c. Lab for Quantum Optics, Shanghai Institure of Optics and Fine Machines,
CAS, Shanghai 201800, P.R.China}

\baselineskip=16pt

\begin{abstract}
We show the possible stable soliton generation for the dark-state
polaritons (DSPs) in an electromagnetic induced transparency (EIT)
medium composed of $\Lambda$-type atoms. Whether the solitons are
dark or bright can be controlled by the coupling field intensity
and the one photon detuning of the probe field. The velocity,
spatial and time widths of the solitons can also be adjusted by
the coupling light.\\

\end{abstract}

\maketitle

\indent The intriguing problem to find new quantum systems whose
wave functions can form solitons and to study their novel
dynamical properties always attracts considerable interests. The
well-known examples include the Ginzburg-Pitaevskii-Gross equation
\cite{1,2} in Bose-Einstein Condensate(BEC) and the Maxwell-Bloch
equation \cite{3} in nonlinear electric medium, where there exists
soliton solutions that respectively describe the properties of
wave functions of the atomic condensate and the electric field. A
natural question then may  be asked: in an interacting system of
atoms and electromagnetic field, if we treat them as a total
system, can the coupled matter-photon system also form a soliton?
An important concrete example may be the light storage in the
electromagnetic induced transparency (EIT)\cite{4} medium which,
in recent years, received much attention due to its potential
applications in the field of quantum information science.

In fact since the technique of resonant enhancement of the index
of refraction without absorption was proposed \cite{5} and many
accompanying striking phenomena was observed \cite{6,7,8,9,10,11},
the light storage with the technique of EIT has been an exciting
research field in current literature, especially after the
"dark-state polaritons" (DSPs) theory was proposed by Fleischhauer
and Lukin et al.\cite{12,13} based on a field theory reformulation
of the adiabatic approximation\cite{14}. DSP is a new quantum
field which is the superpositions of the electric field amplitude
$\varepsilon(z,t)$ and atom coherence $\rho_{cb}$ between two
lower levels of the $\Lambda$ type atoms, and it describes the
total system of the electric field and collective atomic state. In
linear theory where the one-photon detuning of the probe pulse is
zero, the quantum state of the polaritons can be mapped from the
electric field state into the collective atomic excitation without
change when the coupling laser was adiabatically turned
off\cite{12,13}, which implies that DSP is really an elegant
description of the total system.

The paper will discuss the nonlinear properties of the DSPs and
prove that the motion of the DSPs satisfies a (1+1)dimensional
nonlinear Schr\"{o}dinger equation (NLSE) that has possible stable
soliton solutions in an electromagnetic induced transparency (EIT)
medium composed of $\Lambda$ type atoms. Note that the soliton is
not formed by the probe pulse or atom coherence alone, but by the
total state function of them. This is different from the familiar
conventional ones\cite{15} which are formed by the wave function
of a single physical system. The coupling field intensity, along
with the one photon detuning of the probe field, is shown to
decide whether the solitons are dark or bright and other important
parameters, like the time width and the velocity of the solitons.

We consider the quasi 1-dimensional system composed of three-level
$\Lambda$-type configuration atoms with energy levels assumed to
be $E_a>E_c>E_b$. A coherent probe field with positive frequency
part of the electric field $E_p^{(+)}(z,t)$ couples the transition
between the ground state $\lvec{b}$ and the excited state
$\lvec{a}$. $\Delta=\omega_{ab}-\omega$ is the small one-photon
detuning between the carrier frequency $\omega$ and the atomic
transition frequency $\omega_{ab}$. The stable state $\lvec{c}$ is
coupled to $\lvec{a}$ via a coherent coupling field with
Rabi-frequency $\Omega$. We assume the coupling pulse is much
stronger than the probe one and its frequency is
$\nu=\omega_{ac}$. Due to the small intensity of the probe field,
we expand the density matrix in the following form:
\begin{eqnarray}\label{eqn:1}
\rho_{\mu\nu}=\rho^{(0)}_{\mu\nu}+\rho^{(1)}_{\mu\nu}+
\rho^{(2)}_{\mu\nu}+\rho^{(3)}_{\mu\nu}+...
\end{eqnarray}
where $\mu,\nu=a,b$. The quantities $\rho^{(1)}_{\mu\nu}$ is of
the same order of smallness as intensity of the probe pulse, the
$\rho^{(2)}_{\mu\nu}$ is of the second order of the smallness, and
so on. To analyze the nonlinearity of the susceptibility of the
probe light, we calculate the density matrix element $\rho_{ab}$
to the third order. Together with the relation:
$\wp_{ab}n_a(\rho^{(1)}_{ab}+\rho^{(3)}_{ab})=\epsilon_0(\chi^{(1)}(\omega)
+\chi^{(3)}(\omega)|\varepsilon|^2)\varepsilon$, where
$\varepsilon$ is the dimensionless slowly varying amplitude of the
probe pulse $E_p(z,t)$ and $n_a$ being the atom density, the
susceptibility
($\chi\approx\chi^{(1)}+3\chi^{(3)}|\varepsilon|^2$) is then given
by
\begin{eqnarray}\label{eqn:2}
\chi(\omega)\approx-\frac{2g^2N\Delta}{\Omega^2\omega}-
\frac{6g^4N\Delta|\varepsilon|^2}{\Omega^4(1+\frac{-\Delta^2+i\gamma_{ab}
\Delta}{\Omega^2})\omega}
\end{eqnarray}
where $\gamma_{\mu\nu}$ ($\mu,\nu=a,b,c$) is the transverse decay
rate between levels $\mu$ and $\nu$ and here they satisfy:
$\gamma_{cb},\gamma_{ca}\ll\Omega$;
$g=\wp_{ab}\sqrt{\frac{\omega}{2\hbar\epsilon_0V}}$, and N is the
effective total atoms in the quantum volume V. The susceptibility
is related to the refractive index through
$n=\sqrt{1+\chi_(\omega)}$, then the wave vector of the probe
pulse can be approximately calculated by
\begin{eqnarray}\label{eqn:3}
k\approx
k_{ab}+\frac{3}{2}k_{ab}\chi^{(3)}|\varepsilon|^2+\beta_1(\omega-\omega_{ab})+\beta_2(\omega-\omega_{ab})^2
\end{eqnarray}
where
$\beta_1=\frac{\partial{k}}{\partial{\omega}}|_{\omega=\omega_{ab}}=1/(c\cos^2\theta(t))$
is the inverse of the group velocity of the probe pulse,
$\beta_2=\frac{1}{2}\frac{\partial^2{k}}{\partial^2{\omega}}|_{\omega=\omega_{ab}}=
-\tan^4\theta/(\omega_0 c)$ and $k_{ab}=\omega_{ab}/c$. The mixing
angle $\theta$ is defined via $\tan\theta(t)=g\sqrt{N}/\Omega(t)$,
which is a sufficiently slowly time-dependent function. Assuming
that the wave vector of the probe pulse has a narrow spreading
around the central value $k_0$, from the formula (3) the frequency
of the probe pulse can be expanded around $k_0$ in the following
form:
\begin{eqnarray}\label{eqn:4}
\omega\approx\omega_0-\frac{3}{2}k_0V_g\chi^{(3)}|\varepsilon|^2
+V_g(k-k_0)-\beta_2V^3_g(k-k_0)^2
\end{eqnarray}
where $V_g=c\cos^2\theta$ and $\omega_0=k_0c$ are respectively the
group velocity and central frequency of the probe pulse. So the
dispersion equation of probe field $E_p(z,t)$ can be approximately
given by
\begin{eqnarray}\label{eqn:5}
i\frac{\partial{E_p(z,t)}}{\partial{t}}=&(\omega_0-\frac{3}{2}
k_0V_g\chi^{(3)}|\varepsilon|^2)E_p(z,t) &+V_g(\frac{1}{i}
\frac{\partial}{\partial{z}}-k_0)E_p(z,t)\nonumber
\\&-\beta_2V_g^3
(\frac{1}{i}\frac{\partial}{\partial{z}}-k_0)^2E_p(z,t)
\end{eqnarray}

In above derivation the varying of Rabi frequency $\Omega$ with
time is ignored. In fact, the time-dependent character of $\Omega$
brings a reduction/enhancement of the amplitude $\varepsilon(z,t)$
by a contribution $\frac{\dot{\alpha}}{2(1+\alpha)}\varepsilon$
\cite{13,16}to the motion equation of the probe pulse during the
Raman adiabatic passage, where the group index
$\alpha=g^2N/\Omega^2$ and $\dot\alpha=d\alpha/dt$. Consequently
the propagation equation of the probe pulse in the EIT medium is
\begin{eqnarray}\label{eqn:6}
ik_0(\frac{1}{V_g}\frac{\partial}{\partial{t}}+\frac{\partial}{\partial{z}}+\frac{\dot{\alpha}}{2c})
\varepsilon(z,t)=&k_0\beta_2V^2_g\frac{\partial^2}{\partial{z}^2}\varepsilon(z,t)
&-\eta(\omega_0)|\varepsilon|^2\varepsilon(z,t)
\end{eqnarray}
where $\eta(\omega_0)=\frac{3\omega^2_0\chi^{(3)}}{2c^2}$ and $c$
is the vacuum light speed.

As is known that the propagation of the electromagnetic pulse in
an EIT medium can be easily understood in terms of new quantum
fields, i.e., polariton-like fields, which are superpositions of
the dimensionless electric field amplitude $\varepsilon(z,t)$ and
the atom coherence $\rho_{cb}$. The dark-state polaritons and
bright-state polaritons(BSP) here are defined by\cite{17}
\begin{eqnarray}\label{eqn:7}
\Psi(z,t)=\cos\theta(t)\varepsilon(z,t)-\sin\theta(t)\sqrt{N}\rho_{cb}(z,t)\exp(i(k_0-k_c)z)
\end{eqnarray}
\begin{eqnarray}\label{eqn:8}
\Phi(z,t)=\sin\theta(t)\varepsilon(z,t)+\cos\theta(t)\sqrt{N}\rho_{cb}(z,t)\exp(i(k_0-k_c)z)
\end{eqnarray}
where $k_c$ is the wave-vector of the coupling field in $z$
direction, the mixing angle $\theta$ is a function of time, while
the electric field $\varepsilon(z,t)$ and atom coherence
$\rho_{cb}$ are functions of time and z-axis coordinate. Together
with the formula(\ref{eqn:6}), by a straightforward calculation
one can derive the propagation equation of the DSPs:
\begin{eqnarray}\label{eqn:9}
&ik_0(\frac{1}{V_g}\frac{\partial}{\partial{t}}+\frac{\partial}{\partial{z}})\Psi(z,t)
-k_0\beta_2V^2_g\frac{\partial^2}{\partial{z}^2}\Psi(z,t)\nonumber\\
&=ik_0\frac{1}{V_g}\dot{\theta}\Phi-\frac{ik_0}{\cos^2\theta}\sin\theta(\frac{\partial\tan\theta\varepsilon}{\partial
t} +\sqrt{N}\frac{\partial\widetilde\rho_{cb}}{\partial t})
-ik_0\tan\theta\frac{\partial}{\partial{z}}\Phi\nonumber\\
&+k_0\beta_2V^2_g\tan\theta\frac{\partial^2}{\partial{z}^2}\Phi
-\eta(\omega_0)|\varepsilon|^2\varepsilon(z,t)
\end{eqnarray}
where $\widetilde\rho_{cb}=\rho_{cb}(z,t)\exp(i(k_0-k_c)z)$ and
$\dot\theta=d\theta/dt$. The motion equation of BSPs satisfies
\cite{17}:
\begin{eqnarray}\label{eqn:10}
&\Phi=\frac{\sin\theta}{g^2 N}\Bigl(\frac{\partial}{\partial
t}+\gamma_{ba}+i\Delta\Bigr)\Bigl(\tan\theta\Bigr)\Bigl[(
\frac{\partial}{\partial t}+i\Delta)
(\sin\theta\,\Psi-\cos\theta\,\Phi)+g\sqrt{N}\varepsilon\rho_{ca}\Bigr]+i
\frac{\sin\theta}{g\sqrt{N}} F_{ba}.
\end{eqnarray}
where $F_{ba}$ is Langevin noise term which will be omitted in the
following derivation; and the density matrix
$|\rho_{ca}|\sim|\varepsilon|^2$ and
$|\rho_{ca}|\approx|\frac{g\Delta\varepsilon}{\Omega^2}\rho_{cb}|\ll|\rho_{cb}|$
due to the weak probe field and low atomic excitation. By
introducing the adiabatic parameter
$\epsilon\equiv(g\sqrt{N}T)^{-1}$ with $T$ being a characteristic
time \cite{12,17}, we find in the lowest order adiabatic theory
that,
$|\Phi(z,t)|\sim|\frac{g\varepsilon}{\Omega}\rho_{ca}|\approx\frac{|g\Delta|}{|\Omega|^2}|\frac{g\varepsilon}
{\Omega}|^2|\cos\theta\Psi|$ for the ultraslow light case. It is
clear that when $|\Omega|^2\geq|g\Delta|$ or
$|\Omega|^2\sim|g\Delta|$, one has $|\Phi|\ll|\cos\theta\Psi|$
(since the control field is much stronger than the probe one,
i.e., $|\Omega|^2\gg|g\varepsilon|^2$). The typical values of
these parameters in ultraslow light case are \cite{13,15}
$\Omega=1.0\times10^{6\sim8}s^{-1}$, $\Delta=1.0\times10^7s^{-1}$
$g\sqrt{N}=5.0\times10^{12}s^{-1}$ and $g=2.0\times10^{6}$, from
which we find $\cos\theta\sim10^{-(4\sim5)}$ and
$|\Omega|^2\sim|g\Delta|$ and then $|\Phi|\sim|\frac{g\varepsilon}
{\Omega}|^2|\cos\theta\Psi|\ll|\cos\theta\Psi|$.  Hence one can
approximately obtain
$\varepsilon(z,t)\approx\cos\theta(t)\Psi(z,t)$,
$\sqrt{N}\widetilde\rho_{cb}\approx-\sin\theta(t)\Psi(z,t)$ and
$\Phi\approx0$. Substituting these results into the formula
(\ref{eqn:9}) yields
\begin{eqnarray}\label{eqn:11}
&ik\frac{\partial}{\partial{t}}\Psi(\xi,t)
+\frac{\partial^2}{\partial{\xi}^2}\Psi(\xi,t)=-C_n|\Psi|^2\Psi(\xi,t)
\end{eqnarray}
where $k=k_0/(V_g\sin^4\theta)$, the slowly varying nonlinear
coefficient $C_n=\eta(\omega_0)\cot^2\theta\csc^2\theta$, and
$\xi=z-\int^t_{t_0}{d\tau V_g}$ is the coordinate in the rest
frame of the probe pulse. This is a (1+1)-dimensional nonlinear
Schr\"{o}dinger equation (NLSE), which has a bright (dark) soliton
solution when Re($\eta(\omega_0))>0$ (Re($\eta(\omega_0))<0$). As
is known, the NLSE with slowly varying coefficients can be solved
with perturbation theory \cite{18}. The substitution
$t'=\int^t_0dt/k$,
$\Psi(\xi,t)=\frac{\Psi'(\xi,t')}{\sqrt{|C_n/2|}}$ transforms
Eq.(\ref{eqn:11}) into standard form with perturbation
$i\dot{\theta}\frac{1}{2C_n}\frac{\partial{C_n}}{\partial{\theta}}\Psi$
\cite{18,19}:
\begin{eqnarray}\label{eqn:12}
&i\frac{\partial}{\partial{t'}}\Psi'(\xi,t')
+\frac{\partial^2}{\partial{\xi}^2}\Psi'(\xi,t')+2|\Psi'|^2\Psi'(\xi,t')=i\dot{\theta}P(\Psi')
\end{eqnarray}
where$P(\Psi')=\frac{1}{2C_n}\frac{\partial{C_n}}{\partial{\theta}}\Psi'(\xi,t')$.
In the ultraslow light case $V_g\ll{c}$, we have
$P(\Psi)\approx\tan\theta\Psi$. Then, when $\Delta<0$ and
$\gamma^2_{ab}\ll\Delta^2<\Omega^2$, with a perturbation theory
and one can obtain the fundamental bright soliton: \cite{18,20}
\begin{eqnarray}\label{eqn:13}
\Psi'_b=Asech(\sqrt{2}{A\xi}) \exp(i\int_{t_0}^{t'}{d\tau{2A^2}})
\end{eqnarray}
where $A=A_0\cos\theta(0)/\cos\theta(t)$ and $A_0$ is a constant
related to the initial condition. Then
\begin{eqnarray}\label{eqn:14}
\Psi_b=Msech(\sqrt{|C_n|}{M\xi})
\exp(i\int_{t_0}^t{d\tau{M^2}}|C_n|/k)
\end{eqnarray}
or
\begin{eqnarray}\label{eqn:15}
\Psi_b=Msech({\sqrt{|C_n|}M(z-\int^t_{t_0}V_gd\tau)})\exp(i\int_{t_0}^t{d\tau
M^2|C_n|}/k)
\end{eqnarray}
where $M=\frac{A_0\cos\theta(0)}{|C_n(t)/2|^{1/2}\cos\theta(t)}$
is the maximum amplitude of the soliton. From the formula
(\ref{eqn:13}) and (\ref{eqn:14}) one can easily find the spatial
width of the soliton
$\Delta\xi=ln(2+\sqrt{3})|Nc(\Omega^2-\Delta^2)/(6k_0\Omega^2{M^2(t)}\Delta)|^{1/2}
\cos\theta(t)\propto\Omega(t)$, which can be easily controlled by
the coupling light. When the Rabi frequency is adiabatically
reduced, the spatial width narrows according to
$\Delta\xi(t)=\Delta\xi(0)\Omega(t)/\Omega(0)$. Likewise we obtain
the time width of the soliton
$\tau_{FWHM}=ln(2+\sqrt{3})|N(\Omega^2-\Delta^2)
/(6k_0\Omega^2{M^2(t)}\Delta)\cos^2\theta(t)|^{1/2}\propto1/\Omega(t)$,
which is inversely proportional to the Rabi-frequency of the
coupling light. When $\Omega$ is adiabatically reduced, the time
width broadens according to
$\tau_{FWHM}(t)=\tau_{FWHM1}(0)\Omega(0)/\Omega(t)$. The result
that time and spatial widths can be controlled by the coupling
light is an important observation for the properties of the bright
solitions formed by DSPs. To give a numerical estimate of the
spatial width of the soliton, we assume \cite{13,15} the Rabi
frequency of the coupling pulse
$\Omega=1.0\times10^8s^{-1}$,one-photon detuning
$\Delta=1.0\times10^7s^{-1}$, $g\sqrt{N}=5.0\times10^{12}s^{-1}$,
$N=1.0\times10^{13}$, $A_0\cos\theta(0)\approx0.1$, and
$\omega_0=2.0\times10^{15}s^{-1}$, the spatial width can be then
calculated $\Delta\xi\approx2.63\times10^{-1}mm$. This estimation
indicates that this interesting result may be observable under
present experimental techniques (the length of the EIT medium used
in present experiment is about several centimeters\cite{21}). By
choosing $E_0=M(0)$, $L=1/A_0$, and $\tau=g^2N/(\Omega^2(0)A_0c)$
as units to normalize the amplitude of the DSPs, coordinate $z$
and time $t$, respectively, the amplitude of the fundamental
bright soliton can be plotted in Fig.1.

The derivation of the above result is under the condition
Re($\eta(\omega_0))>0$. However, when the coupling field intensity
is adiabatically reduced to $\Omega^2<\Delta^2$ while even
$\Omega^2\sim|g\Delta|$ with $\Delta<0$, the nonlinear coefficient
$C_n$ becomes a negative. Meanwhile the NLSE supports a dark
soliton solution, the amplitude of which is a hyperbolic tangent
function. The time or spatial width of the dark solitons has the
same form as that of the bright ones. On the other hand, if we
initially set a positive one-photon detuning, i.e. $\Delta>0$, the
DSPs can also form dark solitons (for the case
$\Omega^2>\Delta^2$) and bright ones (for the case
$\Omega^2<\Delta^2$).

It is noteworthy that in our formulation we have ignored the
transverse decay rate $\gamma_{ab}$ which can lead to a
contribution $-i\delta|\Psi'|^2\Psi(z,t)$ to the right-hand side
(RHS) of eq.(12), where the coefficient $\delta\propto$
Im$(C_n)/|C_n|\approx\gamma_{ab}|\Delta|/\bigr[(\Omega^2-\Delta^2)^2
+\gamma_{ab}^2\Delta^2\bigr]^{1/2}$ represents a nonlinear loss
(absorption) and instability for the propagation of NS solitons.
For present purpose, when
$|\Omega^2-\Delta^2|\gg|\gamma_{ab}\Delta|$, the effect of
nonlinear absorption can be ignored. The velocity of the solitons
we discussed above is just the group velocity of the probe pulse
and it can also be conveniently controlled by the coupling pulse.
Another interesting issue is that we may extrapolate the idea in
this paper to other interacting systems such as a passive medium
\cite{22} or a Tonks gas \cite{23}, and we may even introduce a
new technique for quantum memory with solitons formed by total
state function of quantum memories (QMEs) and quantum carriers
(QCAs). Due to the strong stability of solitons, quantum
information may be elegantly stored and released against some
environmental noise or external field disturbances in this
technique.

In conclusion we have shown that the dark-state polaritons in an
EIT medium can form the possible stable solitons, and whether the
solitons are dark or bright is dependent on the coupling field
intensity and the one-photon detuning of the probe field. Note
that the kind of solitons are formed by the total wave function of
two different physical systems, i.e., the electromagnetic field
and collective atomic excitations (spin waves). The velocity, time
and spatial widths of the solitons are also shown to be adjustable
by the coupling light. Of course, these results are derived for a
weak probe field and low atomic excitation, i.e., the atomic
density is approximately be treated as constant, and the more
interesting case with a varying density is not studied at present.
This challenging issue, together with the evolution process from
bright solitons to dark ones, deserves further studies in our
future works.

\bigskip

We thank Zhengxin Liu , Guimei Jiang , Xu Cao and Xin Liu for
their helpful discussions. This work is in part supported by NSF
of China under Grant No.10304020. One of the authors (H. J.) is
also supported by China Postdoctoral Science Fund and Shanghai
Postdoc Research Plan.
%\section{}

%\section{Results}
%\section{Conclusions}
%\indent
%ÕýÎÄ

%%%%%%%%%%%%%%%%%%%%%%%%%%%%%%%%%%%%%%%%%%%%%

%%%%%%%%%%%%%%%%%%%%%%%%%%%%%%%%%%%%%%%%%%%%%
\noindent

\newpage
\begin{figure}[ht]
\includegraphics[width=0.8\columnwidth]{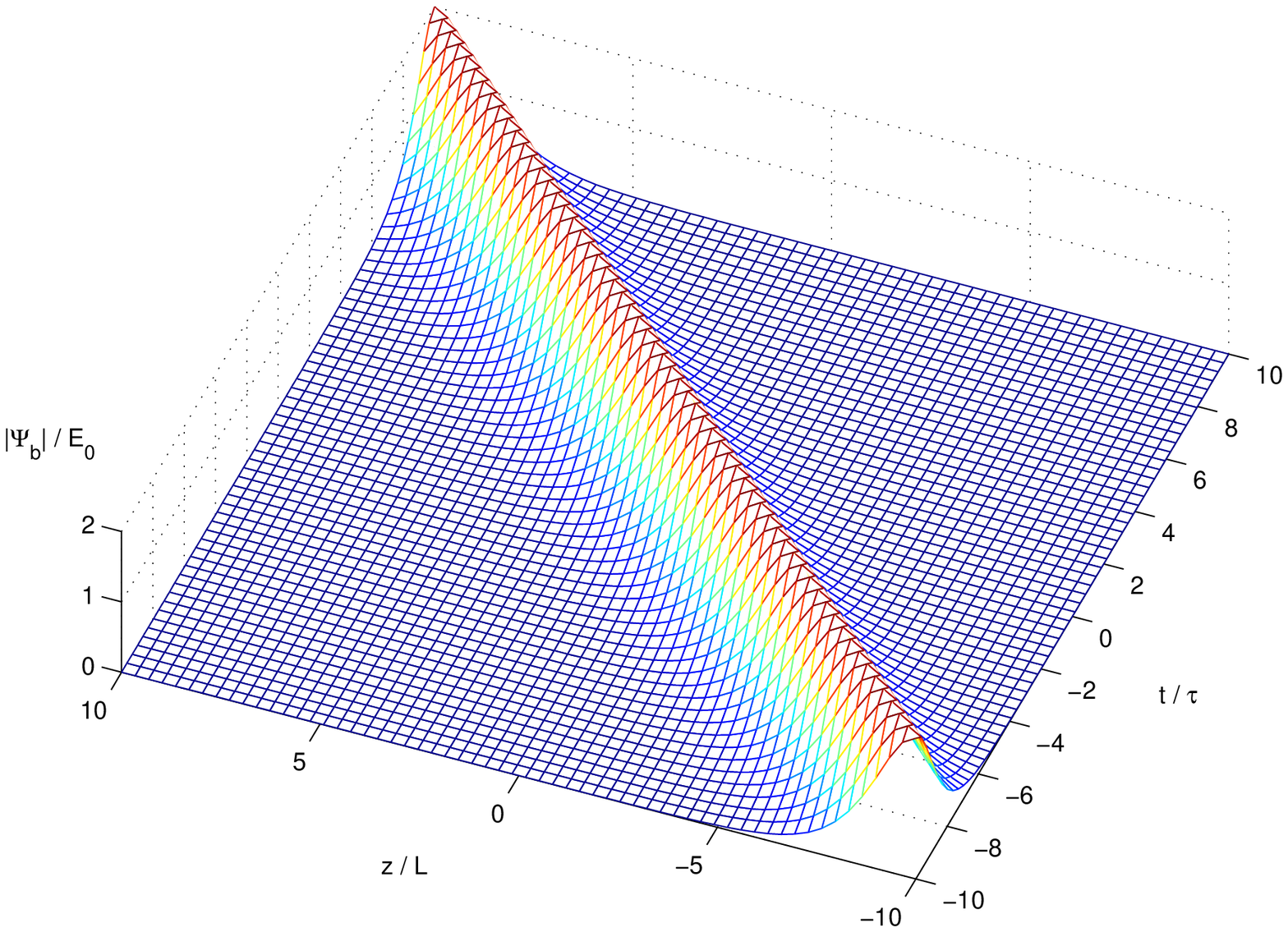}
\caption{The amplitude of the fundamental bright soliton formed by
the DSPs with dimensionless variables. The normalized factors are
$E_0=M(0)$, $L=1/A_0$, and $\tau=g^2N/(\Omega^2(0)A_0c)$ that
represent the units of the amplitude of the DSPs, coordinate $z$
and time $t$.}\label{}

\end{figure}
\end{document}